\documentclass[journal]{IEEEtran}
\usepackage{ifpdf}

\usepackage{amsmath}
\usepackage{algorithm}
\usepackage[noend]{algpseudocode}

\makeatletter
\def\BState{\State\hskip-\ALG@thistlm}
\makeatother

\usepackage{cite}

\ifCLASSINFOpdf
  \usepackage[pdftex]{graphicx}
   \DeclareGraphicsExtensions{.pdf,.jpeg,.png}
\else
\fi
\usepackage{lipsum}

\usepackage{amsmath}
\usepackage{amssymb}
\usepackage{flexisym}
\usepackage{array}
\newcolumntype{C}[1]{>{\centering\let\newline\\\arraybackslash\hspace{0pt}}m{#1}}

\usepackage{color}
\definecolor{dblue}{rgb}{0,0,0.8}

\usepackage{xcolor}

\usepackage{multirow}

\usepackage{subfig}
\usepackage{graphicx}
\usepackage{fixltx2e}

\usepackage{stfloats}
\usepackage{nomencl}

\usepackage{setspace}
\makenomenclature

\begin{document}
%



\title{Exploiting the Benefits of P2P Energy Exchanges in Resilience Enhancement of Distribution Networks }



%

\author{\IEEEauthorblockN{Hamed Haggi and Wei Sun} \\
\IEEEauthorblockA{Florida Solar Energy Center, University of Central Florida, Cocoa, FL, USA\\ Department of Electrical and Computer Engineering, University of Central Florida, Orlando, FL, USA}}

%
%


\maketitle

\begin{abstract}

As the adoption of distributed energy resources grows, power systems are becoming increasingly complex and vulnerable to disruptions, such as natural disasters and cyber-physical threats. Peer-to-peer (P2P) energy markets offer a practical solution to enhance reliability and resilience during power outages while providing monetary and technical benefits to prosumers and consumers. This paper explores the advantages of P2P energy exchanges in active distribution networks using a double auction mechanism, focusing on improving system resilience during outages. Two pricing mechanisms—distribution locational marginal price (DLMP) and average price mechanism—are used to complement each other in facilitating efficient energy exchange. DLMP serves as a price signal that reflects network conditions and acts as an upper bound for bidding in the P2P market. Meanwhile, prosumers and consumers submit bids in the market and agree on energy transactions based on average transaction prices, ensuring fast matching and fair settlements. Simulation results indicate that during emergency operation modes, DLMP prices increase, leading to higher average transaction prices. Prosumers benefit from increased market clearing prices, while consumers experience uninterrupted electricity supply. 
\end{abstract}

\vspace{2mm}

\IEEEpeerreviewmaketitle

\section{Introduction}

\IEEEPARstart{E}{lectrical} power systems are the backbone of today's modern society, supporting critical infrastructures such as communication, transportation, etc. Their secure and reliable operation is vital for societal stability and economic growth. However, in recent years, extreme events, such as cyber-physical-human threats and natural disasters, have posed significant challenges to power system resilience. These threats can significantly affect the operation of power systems, particularly distribution networks (DNs), which are often more vulnerable as disruptions at this level can lead to widespread service interruptions and cascading failures across these systems. Addressing these challenges requires innovative strategies to improve the flexibility, robustness, and adaptability of DNs, ensuring their ability to maintain critical operations during extreme events \cite{haggi2019review}.

In recent years, with the introduction of peer-to-peer (P2P) energy markets, energy trading within distribution networks has become more attractive to customers and even utilities due to its direct and indirect benefits. These markets enable customers to trade energy directly, offering monetary benefits by allowing prosumers to sell surplus energy or purchase cheaper energy from this market instead of the power grid. This financial incentive encourages participation, promoting localized energy management. Beyond economic advantages, P2P energy exchanges enhance reliability and resilience by optimizing resource allocation and reducing dependency on the centralized grid, particularly during disruptions. These markets improve energy availability and flexibility, making distribution networks more adaptive to high-impact events while fostering decentralized, robust energy systems. 

There are many research efforts on the resilience enhancement of DNs using various smart grid technologies which were reviewed thoroughly in \cite{xu2024resilience} with a focus on resilience against natural disasters and climate risks, and \cite{liu2024enhancing}, \cite{rahman2022challenges} and \cite{shibu2024optimising} which focused on cyber resilience. Since the scope of this paper is to explore the benefits of P2P energy exchanges on resilience improvement within distribution networks, we only focus on literature that explored both P2P and resilience improvement (due to space limitation). For instance, a stochastic optimization framework for resilient operation scheduling of interconnected energy hubs, leveraging P2P energy sharings and energy storage to address disruptions like windstorms was introduced in \cite{rezaei2022stochastic}. The framework accounted for uncertainties in renewable energy sources and disturbance durations, optimizing resource allocation for profit maximization under normal conditions and minimizing load shedding in resilient modes. Authors of \cite{arsoon2020peer} introduced a P2P energy bartering framework that enhances the resilience of networked microgrids during extreme events. By enabling decentralized energy sharing through consensus algorithms and proportional sharing, the framework eliminated the need for financial intermediaries and complex regulations. This paper \cite{spiliopoulos2022peer} developed a P2P energy exchange framework to enhance the economic and resilient operation of microgrids. The framework demonstrated significant benefits for system operators and end-users, including economic gains, improved resilience, and reduced carbon emissions, tested across diverse geographical locations and fault scenarios. In \cite{babu2024resilient}, the authors evaluated resilience enhancement in microgrids through P2P energy sharing using a novel resilience metric based on the percolation threshold. Authors of \cite{sadeghi2024peer} leveraged blockchain's P2P trading feature to enhance energy resilience by enabling decentralized and transparent energy transactions. It introduced a mathematical decision-making model to manage P2P energy trading, supported by the theory of planned behavior for demand prediction. 

However, the aforementioned research works did not consider the physical constraints of distribution networks. Also fail to treat consumers as price-maker entities, limiting their role in setting prices. Additionally, economic and market challenges, like fair pricing and balancing incentives, are often overlooked, making it harder to implement these systems effectively for resilience-oriented purposes. These gaps highlight the need for innovative frameworks that address these shortcomings to fully realize the techno-economic potential of P2P energy exchanges in enhancing DN resilience. To this end, the primary contribution of this paper is to develop and implement a resilience-oriented framework to exploit the benefits of network-constrained double-auction-based P2P energy exchanges to enhance power distribution system resilience. Two pricing mechanisms—DLMP and average price mechanism (APM)—are used to complement each other in facilitating efficient energy exchange in both normal and emergency operation modes. In addition, two main real-world scenarios were designed with a focus on the resilience index of the system as well as market signals during the survivability phase.

The rest of the paper is organized as follows: Section \ref{framework} will discuss the P2P framework for resilience enhancement. Section \ref{formulation} and section \ref{Results} will present problem setup, simulation results and numerical analysis, respectively. Finally, section \ref{conclusion} concludes the paper. 

\section{Proposed Resilience Enhancement Framework with P2P Energy Sharing}\label{framework}

\begin{figure}
\centering
\footnotesize
\captionsetup{singlelinecheck=false,font={footnotesize}}
	\includegraphics[width=3.4in,height=2.2in]{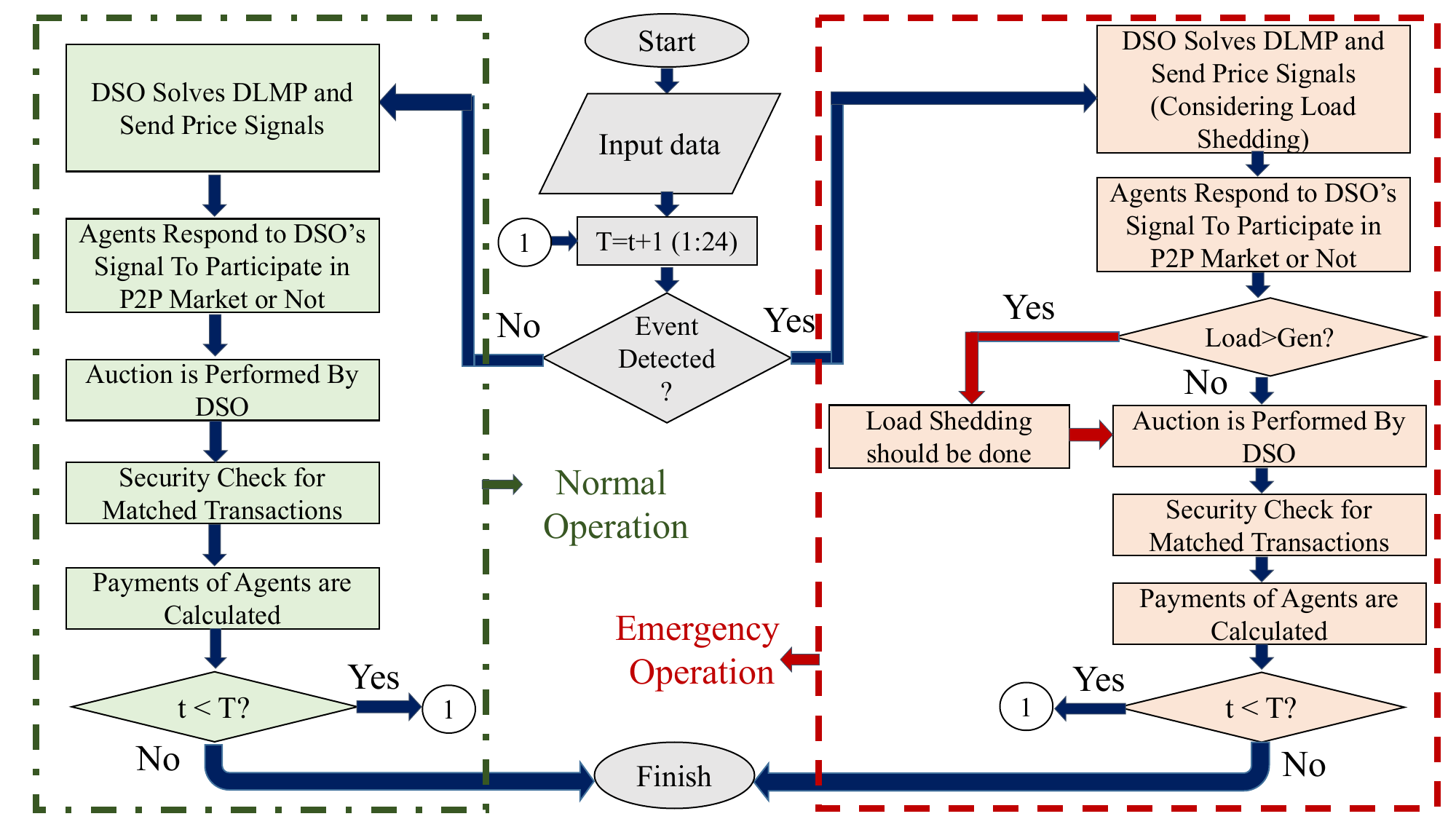}
	\caption{Resilience-oriented double auction based P2P energy exchange algorithm.}
    \label{ResilienceAlgorithm}
\end{figure}

The proposed framework, which is presented in Figure \ref{ResilienceAlgorithm} begins with the DSO collecting bid and ask prices from consumers and prosumers, initiated by sending feed-in tariff (FIT) and DLMP price signals. FIT incentivizes prosumers to trade energy even if unmatched in the P2P market, ensuring surplus energy supports grid resilience. Using a hierarchical double auction mechanism, transactions are prioritized locally at the nodal and zonal levels to minimize grid dependence, reduce losses, and enhance system reliability. The auction clears transactions based on average pricing mechaism, ensuring fairness and strong budget balance. The DSO verifies all trades for compliance with network constraints like voltage and line limits, blocking risky transactions using sensitivity factors. Finally, the payments for consumers and prosumers will be calculated for every successful matching and the matching process is ended. In emergency operation mode (survivability phase), like equipment failure or technical outages, the system operates in islanded mode, relying only on utility-operated DGs and surplus energy from DERs. If demand exceeds supply, load shedding is necessary. The same steps as in normal operation are followed for calculating DLMP and bidding. More information regarding normal operation, DLMP calculation, matching algorithms can be found in our previous work \cite{haggi2021multi}.

\section{Problem Setup}\label{formulation}
A radial distribution network with a set of nodes and edges is defined as $N$ and $E$ respectively. The graph is a tree and the node 0 is the root node. To represent the P2P energy exchange, $N_s$ and $N_b$ are the nodes with generation units representing prosumers and consumers, respectively. The following sections will focus on brief explanations of DLMP and P2P formulation-based on double auction mechanisms with average price mechanism. 

\subsection{Distributional Locational Marginal Price Signal}
DLMP signals in active distribution networks refer to a pricing mechanism that calculates the cost of delivering an additional unit of electricity at specific nodes within the distribution network. Unlike traditional pricing, DLMP accounts for real-time factors such as network constraints, losses, congestion, and the operation of distributed energy resources. In active distribution networks—where energy flows can be bidirectional due to these distributed resources—DLMP provides precise price signals that encourage efficient energy production and consumption. 

\begin{equation}\label{objective function}
   min  \sum_{t=1}^{T}\left [ \sum_{i=1}^{N} (C^g_{i,t} \times P^g_{i,t}) + (C^g_{0,t} \times P^g_{0,t}) + (VOLL\;. P^s_{i,t}) \right ]\\
\end{equation}

Where the first and second term refer to generation cost and last term refer to load curtailment cost. Due to page limitations, the following constraints' formulation can be found in \cite{papavasiliou2017analysis}, voltage constraints and limits, line flow constraints with and without loss, branch current constraint, and generation limit. The only constraint that is updated in this reference \cite{papavasiliou2017analysis} formulation is nodal power balance which now due to emergency situation includes active and reactive load-shedding variable as well as their appropriate upper and lower bound. 

Finally, based on duality analysis of the above SOCP problem formulation and KKT conditions, nodal marginal prices can be obtained from (14).
\begin{equation}
\pi_i=\omega_1 . \pi_{A_{i}}+ \omega_2. \mu_i+\omega_3 . \mu_{A_{i}}+\omega_4. \eta_i +\omega_5.\eta_{A_{i}}
\end{equation}
Where $\pi, \mu, \eta$ are active power price, reactive power price, and contribution of complex power at node $i$. Note that index $A_i$ refers to the ancestor node of $i$. More information about dual variables calculation, $\omega_i$, and decomposition of DLMP can be found in \cite{papavasiliou2017analysis}.

\subsection{Entities in Multi-Round Double Auction}
In a double auction setting, the primary participants are sellers (prosumers), buyers (consumers), and the auctioneer, typically the DSO. Prosumers are indexed by $n$, and consumers are indexed by $m$.
Prosumers specify their asking price $Sp$ and the quantity of power $Sq$ they are willing to trade in the P2P market. The payoff for a prosumer excluding additional costs, is calculated as $SW_n$:

\begin{equation}\label{prosumer welfare}
SW_n = \left\{\begin{matrix}
 \sum_{m=1}^{N_b}\left ( \gamma _{nm} - Sp_n\right )\times Sq_{nm},&if\ pros.\ n \ wins \\ 
0 \qquad\qquad\qquad\qquad\qquad & otherwise.
\end{matrix}\right.
\end{equation}

Where the cleared auction price for the successful transaction between prosumer $n$ and consumer $m$ is $\gamma _{nm}$. On the other hand, Consumers specify their bid price $Bp$ and the amount of energy needed $Bq$, respectively. they wish to purchase in the P2P market. The payoff for consumer $m$ from auction xcluding any additional costs, is defined as, $BW_m$, is defined as: 
\begin{equation}\label{consumer welfare}
BW_m = \left\{\begin{matrix}
 \sum_{n=1}^{N_s}\left ( Bp_m - \gamma _{mn} \right )\times Bq_{mn},&if\ con.\ m \ wins \\ 
0 \qquad\qquad\qquad\qquad\qquad& otherwise.
\end{matrix}\right.
\end{equation}
Finally, the total social welfare for all prosumers and consumers is defined as $TW$ which is the sum of  $SW_n$ and $BW_m$:
 \begin{equation}\label{total welfare}
 \quad TW= \sum_{n=1}^{N_s}SW_n \ + \sum_{m=1}^{N_b}BW_m
\end{equation}

There are significant challenges in directly applying existing research on prosumers and consumers from wholesale or retail markets to distribution networks, as detailed in Section I. To address these challenges, we propose a hierarchical auction mechanism called Multi-Round Double Auction with an Average Price Mechanism (MRDA-APM). This auction consists of three rounds: nodal, zonal, and distribution network (DN)-level auctions. The APM ensures a strong budget balance by equally distributing all benefits among prosumers and consumers, preventing the auctioneer (DSO) from profiting. We consider 15-minute intervals for P2P energy exchanges, though the interval length can vary based on system size, agent numbers, and auction design.
After collecting bids and offers from agents, the DSO arranges prosumers' ask prices in ascending order and consumers' bid prices in descending order:
\begin{equation}
Sp_1< Sp_2< ...< Sp_n, \quad \forall n\in S
\end{equation}
\vspace{-0.3cm}
\begin{equation}
Bp_1> Bp_2> ...> Bp_m, \quad \forall m\in B
\end{equation}
The average transaction price between prosumers selling price and consumers' buying price is calculated as $\gamma$: 
\begin{equation}
\gamma = \frac{\sum_{n=1}^{N_s} Sp_n+\sum_{m=1}^{N_b}Bp_m}{N_s+N_b}
\end{equation}
A prosumer wins if and only if $Sp_n< \gamma$ and From the consumer's perspective, $Bp_m> \lambda$. The MRDA-APM allows agents to negotiate first at the nodal level, where transaction costs are minimal due to power loss, congestion, and voltage regulation costs. If a prosumer fails to find a match, they proceed to zonal and then DN-level auctions. This multi-layer process ensures all agents have the opportunity to participate, preventing any single agent from dominating the market within a time interval. At each matching stage, the clearing price is the average of the prosumer's ask price and the consumer's bid price, as per the APM. This ensures that all trading benefits are shared between the matched peers rather than accruing to the DSO. While the DSO does not profit directly, it benefits indirectly by reducing the need to purchase energy from the upper-level network, thanks to increased participation from small-scale prosumers. Finally, If the auction concludes with unmatched prosumers, they sell their remaining energy to the grid at the Feed-in Tariff (FIT) price. Unmatched consumers purchase any additional required energy from the grid at the DLMP price. For more information please refer to our previous work \cite{haggi2021multi}, Algorithms 1 and 2. After the matching process, to ensure the network operates smoothly without issues like voltage problems or congestion, the DSO needs to monitor the energy trades between peers to avoid any violation. By using sensitivity factors, the DSO can check if a specific transaction might cause any technical issues in the network. After final approval of DSO regarding transactions, final payments will be sent to the account of both prosumers and consumers regarding each successful transaction. Interested readers are encouraged to check our previous work regarding the details of algorithms, sensitivity factors, etc.

\subsection{Calculation of Resilience Index}
The performance of the proposed framework should be evaluated using an appropriate resilience index. However, there is not any consensus metric for power system resilience evaluation. Therefore, this paper consider the vertical axes of resilience curve related to load served and load shedding. To this end, resilience index can be defined as below:
\begin{equation}
RI_t=\left [ 1-\left ( \frac{P^s_{i,t}}{P^{ld}_{i,t}} \right ) \right ] \times 100, \qquad \forall t\in T
\end{equation}
where $P^s_{i,t}$ and $P^{ld}_{i,t}$ represent load shed and total load,.

\section{Simulation Results and Numerical Analysis}\label{Results}
The implementation of the proposed P2P framework aimed at improving system resilience was executed on a modified 33-node distribution test system, incorporating pre-defined zones for improved analysis. More information regarding how these zones are chosen can be found in our previous work \cite{haggi2021multi}. Simulation were conducted on a PC equipped with an Intel Core i7 processor and 16 GB of RAM. The simulations were carried out on the General Algebraic Modeling System (GAMS) and MATLAB software platforms \cite{ferris1999matlab}, ensuring efficient execution and comprehensive data processing.

\subsection{Simulation Results and Analysis}
\begin{figure}
\centering
\footnotesize
\captionsetup{singlelinecheck=false,font={footnotesize}}
	\includegraphics[width=3.5in,height=1.9in]{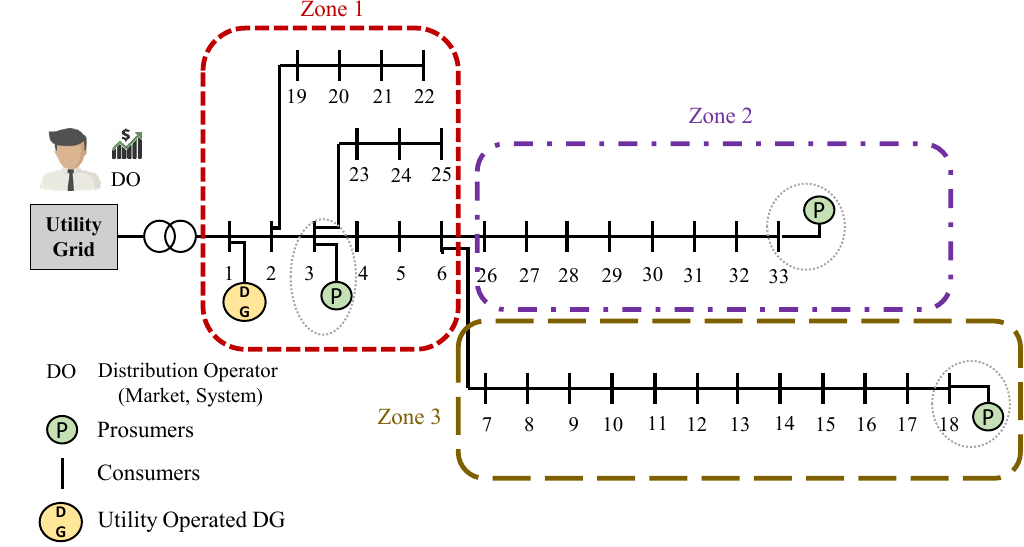}
	\caption{Modified 33-node test system.}
    \label{testsystem}
\end{figure}

The modified 33-node test system, as widely discussed in \cite{baran1989network}, is shown in Figure \ref{testsystem}. A 2.5 MW utility-operated distributed generation (DG) is installed at node 1 with an operating cost of 50 \$/MWh. Three 1000 kW photovoltaic (PV) systems are distributed across nodes 3, 18, and 33, representing the three prosumers. Notably, these nodes serve as both producers and consumers due to their combined energy generation and consumption capabilities. The network of 33 nodes is clustered into three distinct zones, facilitating a comprehensive analysis of system performance within varied operational contexts. Hourly load data for the system is acquired from \cite{bai2017distribution}, while the PV data for the first day of July 2019 is sourced from the California Independent System Operator. Furthermore, the hourly locational marginal prices (LMPs) for the upstream network used in the study are obtained from the Pennsylvania–New Jersey–Maryland Interconnection (PJM) \cite{bai2017distribution}. In this paper, all simulations are conducted with a time interval of $\Delta t$ = 15 minutes, enabling a comprehensive assessment over a complete day. In the context of emergency scenarios, it is assumed that the transformer/tie line connecting node 1 to the upstream grid is out of service due to either a mechanical failure or a physical attack. Consequently, the system is forced to transition into an emergency operation mode, thereby isolating the test system from the upstream grid.

\subsection{Resilience Analysis with and without P2P}
This study assumes an outage/failure on the transformer/tie line connecting the distribution network to the upstream network. The results for the emergency situation are predicated on a one-day repair period, following which the system transitions from an islanded state back to a grid-connected mode. During emergency operations, the grid operates in an isolated mode, unable to procure power from upper levels. The only source of supply is the utility-operated DG, capable of generating a maximum of 2.5 MW. Consequently, during periods when the electricity demand surpasses the generation, the DSO is forced to curtail some load. 

\begin{figure}
\centering
\footnotesize
\captionsetup{singlelinecheck=false,font={footnotesize}}
	\includegraphics[width=3.5in,height=1.9in]{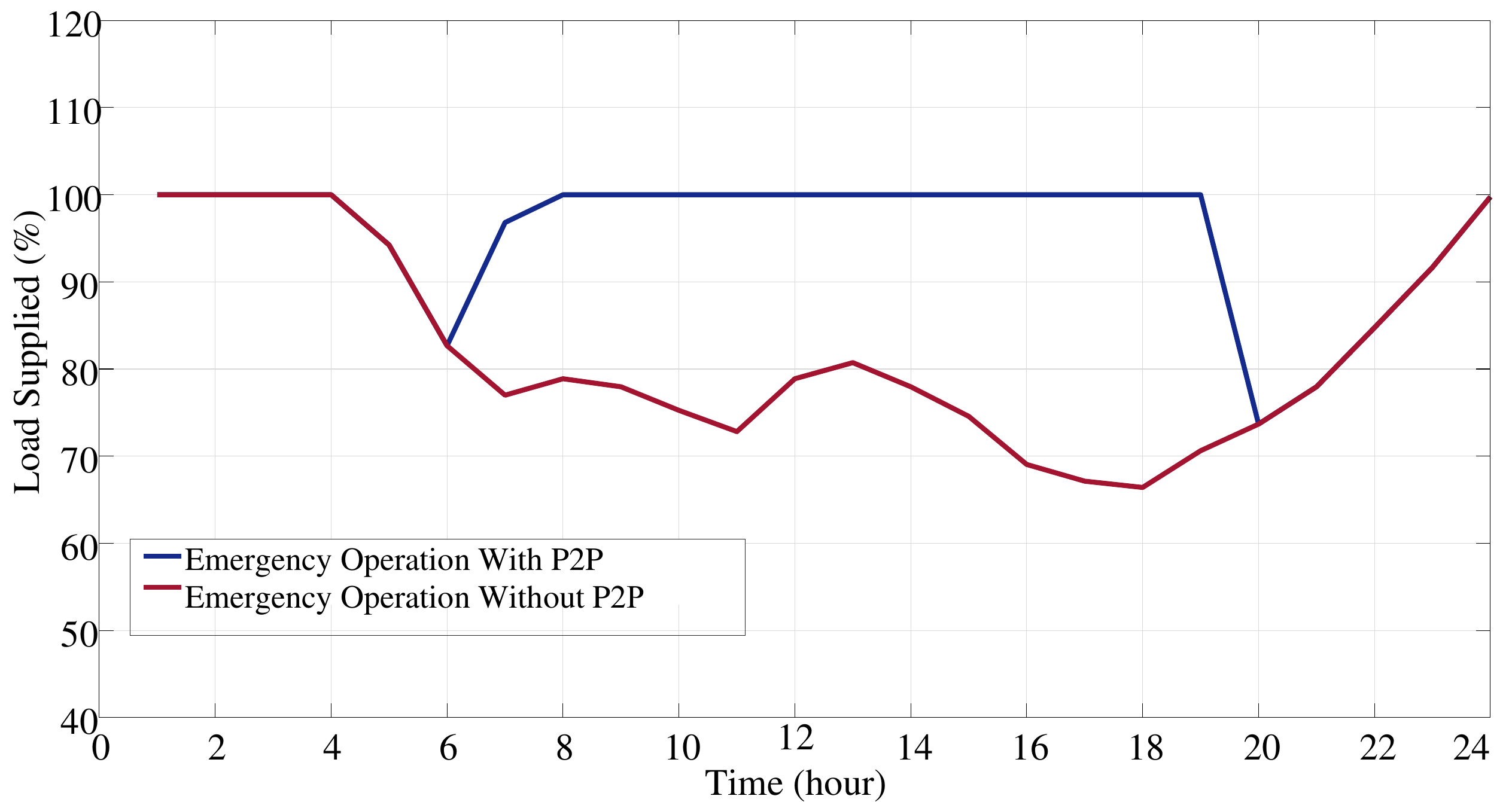}
	\caption{Served Load with and without P2P.}
    \label{LSe}
\end{figure}

Served load results with and without P2P energy exchanges are shown in  Fig. \ref{LSe}. Notably, between the hours of 8 am and 7 pm, the surplus power generated by the prosumers is effectively shared through the P2P mechanism, resulting in the complete fulfillment of the load, as denoted by 100\% load supply (resilience index of 100\%). In contrast, without considering P2P energy exchanges, the system experiences instances where the load cannot be entirely met. During the hours of 8 am and 7 pm, the surplus energy generated by the prosumers is insufficient to supplement the utility-operated DG adequately, thereby leading to a partial supply of the overall load demands. These findings underscore the significant role of P2P energy exchanges in optimizing load management and enhancing the overall reliability of the energy distribution system.

\begin{table}[]
\footnotesize 
\captionsetup{font={footnotesize,sc}}
\centering
\caption{Resilience Index (\%) for Cases with and without P2P}
{\begin{tabular*}{21pc}{@{\extracolsep{\fill}}|c|c|c|c|c|c|@{}}
\hline \label{RI}\centering
Case &T=13 &T=14 &T=15 &T=16 &T=17 \\
\hline \hline
Without  &80.7 &77.9  &74.5 &69.0 &67.1 \\
P2P  & & &  & & \\
\hline
With P2P &100 &100 &100  &100 &100 \\
\hline
\end{tabular*}}{}
\end{table}

\begin{figure}
\centering
\footnotesize
\captionsetup{singlelinecheck=false,font={footnotesize}}
	\includegraphics[width=3.5in,height=1.9in]{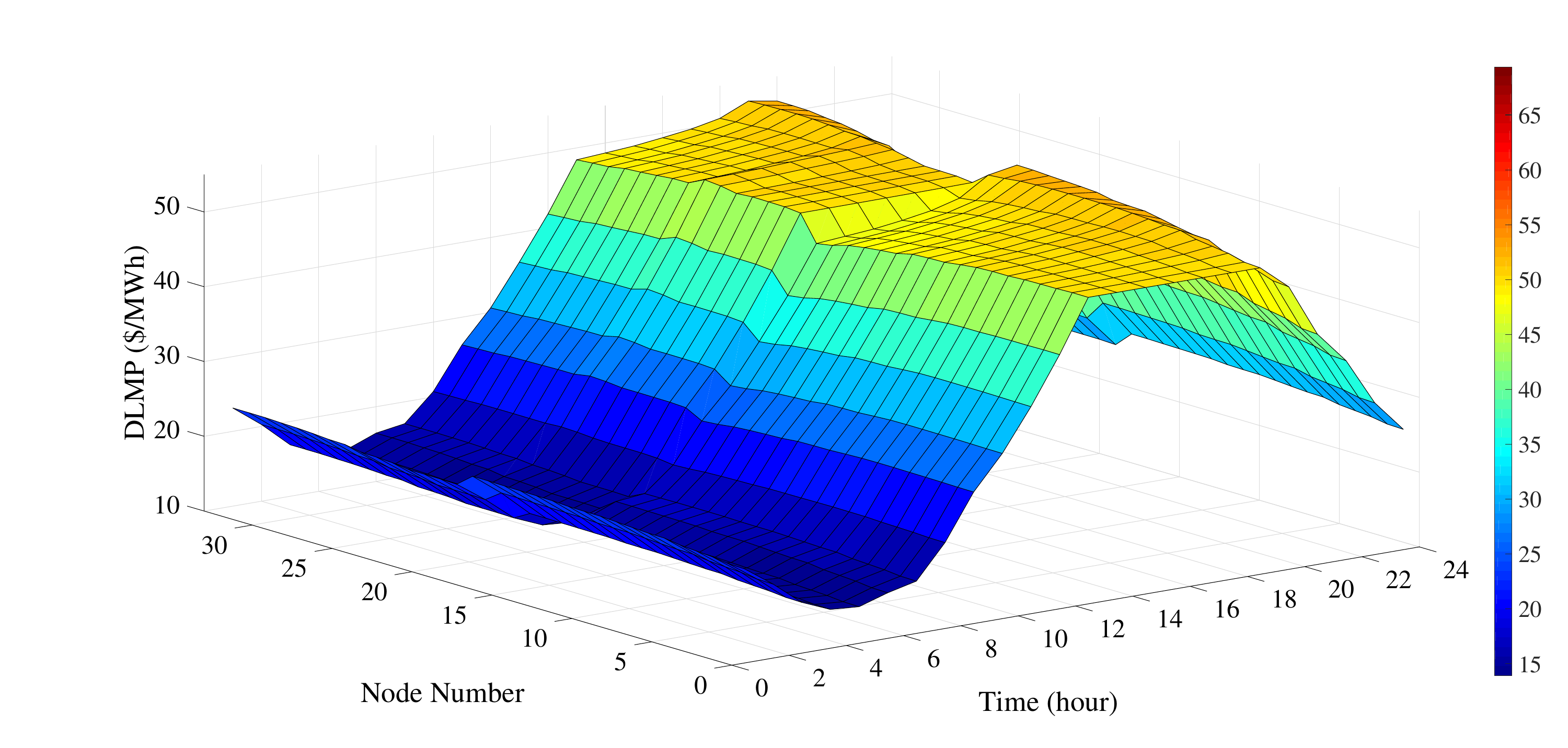}
	\caption{Normal operation DLMP considering P2P.}
    \label{NDLMP}
\end{figure}
\subsection{Normal and Emergency DLMPs with and without P2P}
\begin{figure}
\centering
\footnotesize
\captionsetup{singlelinecheck=false,font={footnotesize}}
	\includegraphics[width=3.5in,height=1.9in]{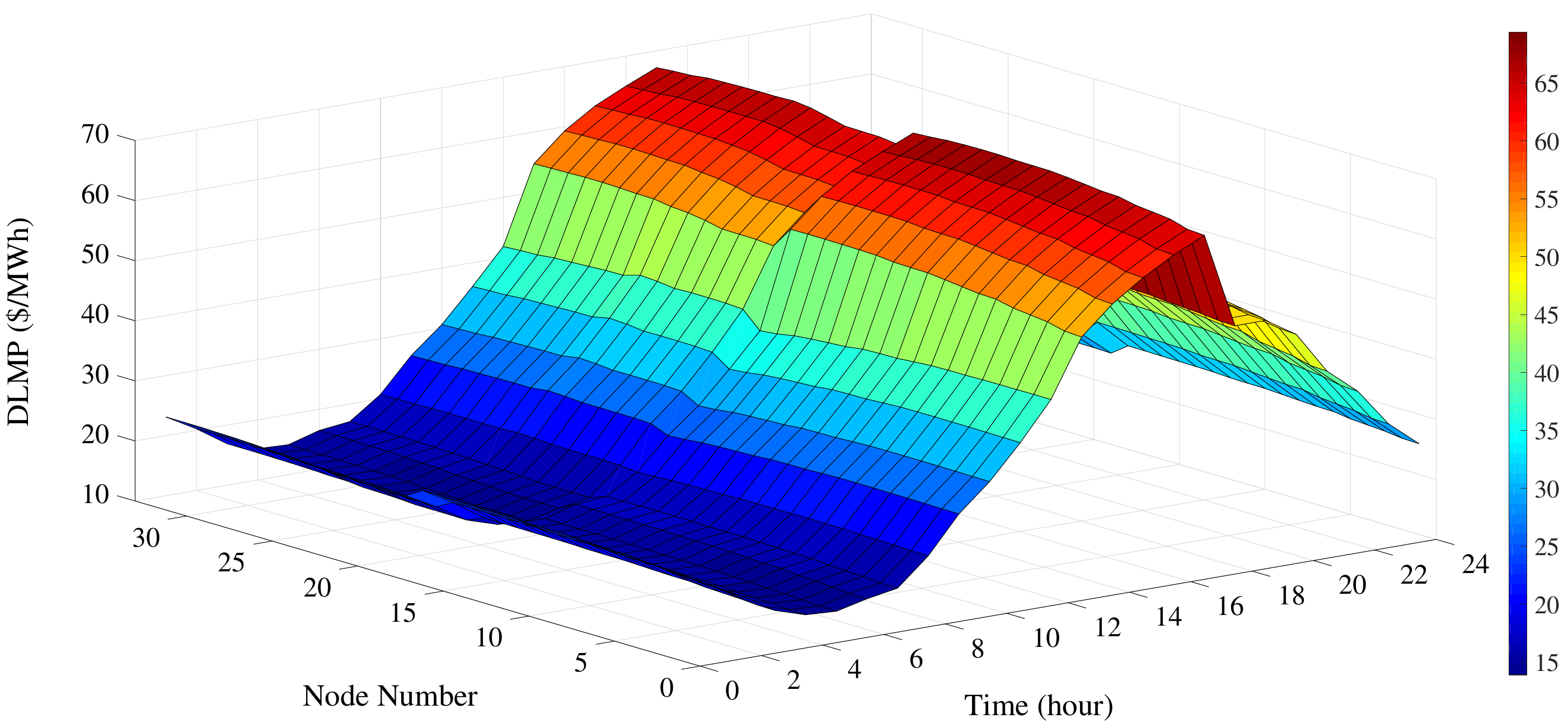}
	\caption{Emergency operation DLMP considering P2P.}
    \label{EDLMP}
\end{figure}

During standard operational conditions, agents within the system are provided with DLMPs as the upper bound of bidding, thereby establishing a reference point for market participation. However, in emergency scenarios, the DLMP exhibits a noticeable increase, owing to the highest bid of prosumers due to power shortage. Consequently, the DLMP during emergency periods surpasses the values observed during normal operating conditions, representing an aberration in price signaling which sheds some light on the fact that during emergencies prosumers prefer to sell their electricity at higher prices to other consumers and benefit from the emergency market. Under these circumstances, prosumers respond by elevating their selling prices, while consumers are forced to increase their bidding amounts to secure successful participation in the auction, ensuring the uninterrupted provision of their energy requirements. The outcomes of this phase are visually represented in Figure \ref{NDLMP}  and Figure \ref{EDLMP}, highlighting the notable disparity between the DLMPs during normal and emergency operation modes. For this analysis, it is assumed that the system is subjected to a disruption at hour 13, with a projected repair duration of 5 hours.\par

A comparative analysis of Figure \ref{NDLMP} and Figure \ref{EDLMP} indicates noticeable differences in DLMP values from hours of 1 pm to 5 pm during normal and emergency operations. Within Figure \ref{EDLMP}, the observed divergence in DLMP values during the emergency phase prompts consumers to adjust their bidding strategies in an effort to secure energy procurement. This bidding adjustment is primarily instigated by the proactive response of prosumers, who opt to sell their surplus energy at the higher price, recognizing the strain on the system's load supply capacity. Consequently, this contributes to an increase in the average P2P transaction prices, as shown in Table \ref{ATP}. The table distinctly illustrates a considerable rise in the average transaction prices (ATPs) during the emergency operation. Moreover, the analysis outlined in Figure \ref{EDLMP} suggests that during the outage period, nodal prices experience a surge, primarily attributed to heightened bidding activities and subsequently elevated clearing prices. The resilience index, as presented in Table \ref{RI}, further underscores the impact of P2P energy sharing on the system's resilience during emergency operations. Notably, the incorporation of P2P energy sharing serves to improve the overall resilience index, thereby enhancing the system's capacity to withstand and recover from disruptive events.
\par

\begin{table}
\footnotesize 
\captionsetup{font={footnotesize,sc}}
\centering
\caption{Normal and Emergency Operation DLMP and P2P Market Average Transaction Prices}

{\begin{tabular*}{21pc}{@{\extracolsep{\fill}}|c|c|c|c|c|c|@{}}
\hline  \label{ATP}
Price  &T=13 &T=14 &T=15 &T=16 &T=17 \\
(\$/MWh) & & &  & & \\
\hline \hline
DLMP  &49.27 &49.43 &49.61  &49.92 &50.23 \\
Normal  & & &  & & \\
\hline
ATP &32.36 &25.88 &27.14  &24.41 &27.57 \\
Normal  & & &  & & \\
\hline
DLMP  &54.64 &59.34 &62.88  &65.59 &67.97 \\
Emergency  & & &  & & \\
\hline
ATP &41.04 &38.72 &43.04  &49.22&43.71 \\
Emergency  & & &  & & \\
\hline
\end{tabular*}}{}
\end{table}

\section{Conclusion}\label{conclusion}

This paper explored the benefits of P2P energy exchanges within power distribution systems for enhancing resilience using a multi-round double auction mechanism. Two pricing schemes—DLMP and average price mechanism—were considered to evaluate the economic performance in both normal and emergency operation modes. Simulation results showed that P2P energy exchanges benefit both prosumers and consumers from two perspectives: (1) during emergency times, prosumers can share their excess energy with consumers, supplying part of their demand and thus increasing resilience; (2) by allowing energy exchange at better prices compared to the grid. During disruptions, prosumers provide their surplus energy to consumers at higher prices than in normal operation modes, which reflects real-world conditions. The numerical findings reveal a marked increase in sellers' revenue during emergency operations. On the other hand, the cost-saving potential for buyers declines compared to the baseline case study due to paying extra money for having electricity during the time that utilities can't provide it.

\setstretch{1}

\bibliographystyle{IEEEtran}

\bibliography{P2P_Resilience}

\end{document}